\documentclass[draftcls,onecolumn,11pt]{IEEEtran}
\usepackage[T1]{fontenc}
\usepackage{graphicx}
\usepackage{textcomp}
\usepackage{pifont}
\usepackage{comment}
\usepackage{physics}
\usepackage{listings}
\usepackage{dsfont}
\usepackage{subfigure}
\usepackage{algorithm,algpseudocode}
\usepackage{amsmath,amssymb,amsfonts}
\usepackage{hyperref}

\begin{document}

\title{Simulation of fidelity in entanglement-based networks with repeater chains}

\author{David Pérez Castro, Ana Fernández Vilas, Manuel Fernández-Veiga, Mateo Blanco Rodríguez  and Rebeca P. Díaz Redondo \thanks{David Pérez Castro, Ana Fernández,  Manuela Fernández Veiga, Mateo Blanco and Rebeca P are with the atlanTTic Research Center (IC Lab), University of Vigo, Spain.
Corresponding author: Ana Fernández Vilas. avilas@uvigo.es}

}

\maketitle

\begin{abstract}
We implement a set of simulation experiments  in NetSquid  specifically  designed for estimating the end-to-end fidelity across a path of quantum  repeaters or 
quantum switches. The switch model includes several generalizations which are not currently
available in other tools, and are useful for gaining  insight into practical and realistic
quantum network engineering problems: an arbitrary number of memory registers at the switches,
simplicity in including entanglement distillation mechanisms, arbitrary switching topologies, 
and routing protocols. An illustrative case study is
presented, namely a comparison in terms of performance between a repeater chain where
repeaters can only swap sequentially, and a single switch equipped with multiple memory
registers, able to handle multiple swapping requests. 

\end{abstract}

\begin{IEEEkeywords}
Quantum networks, entanglement distribution, resource allocation, simulation.
\end{IEEEkeywords}

\section{Introduction}
\label{sec:intro}

Long-distance quantum communications and distributed quantum computing tasks 
necessitate efficient entanglement distribution among the parties to be 
feasible at  practical scale in order to process quantum  information in 
better and more capable  ways than those achievable with classical 
communications~\cite{Azuma2021,Cacciapuoti2020,Rohde2021}. Also,
high-rate entanglement distribution is key for realizing 
information-theoretical secure communications~\cite{Tamaki2018}.  
However,  long-distance distribution of entangled states requires the 
use of quantum repeaters (QR) or quantum switches (QS) as intermediate 
devices to propagate sequentially the link-level entanglement ~\cite{Deutsch1996,Pirandola2017}, so that end-to-end
entangled (EEE) states between two arbitrary nodes are created. 
Functionally, a QR and a QS are similar but QRs are simpler, since they 
are basically two-port devices that extend a linear chain of links, 
while QSs could in principle allow for more general internal quantum 
connections from an input to an output port~\cite{Azuma2023}. 
As in classical networking, QRs would enable the construction of general 
physical interconnection topologies for a quantum network.

Controlling the delay, throughput, and quality (in fidelity, a 
quantitative measure of entanglement) of a quantum network based on 
repeaters or switches is thus a fundamental challenge for practical
applications~\cite{Li2022,Inesta2023}. Different analytical models 
are being developed for elucidating the performance of quantum
networks under a variety of assumptions.  Though the insights and design 
principles gained with analytical models are invaluable to understand the 
complex relationships between physical parameters and network performance,  
the known analytical  tools can only solve simple scenarios, restricted to 
configurations with few nodes, considering only quantum repeaters, or 
specific entanglement protocols, e.g.~\cite{Tillman2024,Andrade2024,Zang2024}.
Furthermore, these works rely on idealized assumptions about the diverse
impairments that quantum devices or quantum channels may exhibit, e.g., loss,
state/memory decoherence, noisy entanglement swapping, or depolarization errors. 

Given the fragility of these basic quantum information processing tasks, 
and the breadth of physical parameters that underlie the behavior of quantum
devices and channels,  tailored quantum-network simulators are often used 
in numerical studies~\cite{Coopmans2021,Simulaqron,Diadamo2021,Wu2021} for 
a twofold reason: (i) to obtain a precise evaluation of non-trivial network
configurations, when the number of physical variables required to model
accurately each QR and the number of intermediate nodes is too large; 
(ii) to test and compare, in a quick experiment, different combinations 
of architectures and protocols suitable for a given computational goal. 
There exist many algorithms, for instance, for distributing entangled
states~\cite{Zang2023}, for using of purification vs. error correction at 
the nodes~\cite{Duer2007}, entanglement swapping 
protocols~\cite{Dahlberg2019}, routing protocols~\cite{Kar2023}, etc.

For the numerical characterization a quantum network, the throughput 
and fidelity of and EEE link are arguably the key performance indicators. 
The throughput is defined as the rate of successful entangled states 
created between two endpoints, and depends on the  physical network 
constraints, the quantum resource constraints (the number of memory 
qubits at the QR), and the embedded protocols. Fidelity is a measure of 
the quality of the generated entangled states, and is significant 
since applications depend critically on a minimum fidelity to be actually 
useful (e.g., quantum key distribution demands fidelity above $0.85$, 
whereas distributed quantum computing needs fidelity beyond $0.9$,
approximately). Therefore, it is of interest to be able to quantify 
accurately both the entanglement rate and the fidelity in complex networks, 
at least to validate whether a particular quantum information processing 
task is possible. Despite the flexibility of simulation tools for this 
purpose, most of the quantum simulators are not focused on
facilitating the estimation of rate and fidelity while simultaneously 
giving precise control on the variables of the main network elements: 
sources, channels, and quantum protocols.

% Contributions
In this paper, we present the design principles, the operating model, 
and the implementation of a set of  experiments in NetSquid to  simulate  advanced routing experiments on entanglement-based quantum networks. We also provide a 
comparison of two network configurations difficult to study analytically. More specifically, our contributions are
the following: First, we design and implement a experimental setting (based on NetSquid~\cite{Coopmans2021}) that simplifies the preparation and running of 
numerical experiments on quantum networks based on entangled resources. Differently to 
other simulation approaches in the literature, which rely on the abstraction  on quantum circuits or ignore 
the real behavior of the devices, our focus is put on the quantum processes, devices, 
and  entities that generate, manipulate, or consume entanglement as a 
fundamental resource. Secondly, our setting adopts a similar protocol as the one described in \cite{zhao2022e2e}, the E2E fidelity aware routing; and  fidelity on a path is estimated in
advance to the creation of the entanglement. Finally, we  conduct two detailed experiments to assess the performance of 
two configurations: a multi-hop network of repeaters, and a two-hop network
wherein QRs have a finite but  arbitrary number of qubits.

% Paper organization
The remainder of this paper is organized as follows. Related works are briefly
reviewed in Section~\ref{sec:related-work}. Section~\ref{sec:background} 
introduces the quantum concepts and mechanisms that are needed for 
subsequently understanding  the system model, presented in 
Section~\ref{sec:Description}. We describe the details of our implementation 
in Section~\ref{sec:implementation}, present the network architectures
in Section~\ref{sec:scenario}, and discuss the numerical results obtained 
with the tool in Section~\ref{sec:results}. The concluding remarks of our 
work are given in Section~\ref{sec:conclusions}.

\section{Related Work}
\label{sec:related-work}

Performance evaluation via mathematical modeling and analysis has been the
focus in prior works for deriving fundamental limits of quantum networks,
develop new insights, and provide a collection of baseline results to use
in numerical or experimental settings. However, this closed-form assessment  
is generally tractable solely for simple scenarios: configurations with 
one or two 
nodes~\cite{Nain2020,Vardoyan2021a,Nain2022,Panigrahy2022,Kamin2023,Tillman2024}, 
linear  chains (i.e., isolated paths in a network)~\cite{Brand2020,Andrade2024}, 
specific entanglement distribution 
protocols~\cite{Shchukin2019,Zangi2023,Zang2024}, or one-shot 
communication instances in which only a single request at a time is
considered~\cite{Vardoyan2020,Vardoyan2021}. At the system level, models 
for performance analysis of quantum networks addressing queue stability, 
control, or scheduling policies  have also been 
investigated~\cite{Zhao2021,Vasantam2021,Vardoyan2022,Vasantam2022,Vardoyan2023}, 
mainly under idealized assumptions on the diverse impairments that 
quantum devices or quantum channels may exhibit, e.g., loss, state/memory
decoherence, noisy entanglement  swapping, or depolarization errors.

Discrete-event simulation is used in some of these works for validating 
the assumptions of the formal model, like in~\cite{Vardoyan2021} or
\cite{Coopmans2021}, \cite{koz2020designing}, which exploit NetSquid to
corroborate the results predicted by a Markovian analysis of a single
switch and a single repeater. Different quantum memory technologies and 
simulator configurations were compared to evaluate general performance 
of the system, but in either case a generalized network architecture
remained unimplemented, specially the network size for the switch case, 
and the quantum memory sizes in the 2-switch case. 
%Moreover, we consider a lack of a framework to properly evaluate time as a variable, not accounting for protocols and other processes  present in a network.
Alternatively, numerical simulation forms the basis for studying the 
theoretical limits of quantum switches in works like~\cite{Tillman2024}, 
where the capacity region under different conditions is completely bounded
through convex analysis. Nevertheless, numerical simulation has a high
computational load, and the analysis here deals with a single switch as 
well. Another limitation of current simulation experimental settings is that they have not been optimized or conceived for implementing and evaluating
key metrics like fidelity as a resource, yet fidelity is fundamental for routing, 
distributed quantum computing and quantum 
communication~\cite{Li2022,Goodenough2024,Kamin2023}.  

Our work sets the ground base of a simulation framework designed for
experimentation on arbitrary network graphs, based on a more general operating
model supporting quantum switches and routing, but with the capability of precise
configuration of the physical properties of channels, sources, quantum memories, and quantum 
programs.

\section{Entanglement generation and distribution}
\label{sec:background}

At a fundamental level,  a quantum network can be conceptualized as a 
system for the efficient distribution of high-quality entangled states over
long distances and for an arbitrary group of participants. This Section 
briefly reviews the mechanisms necessary for the generation of entanglement 
over multiple links in a network, and for enhancing or preserving the qubits 
as close as possible to a state with maximal entanglement.  

\subsection{Entanglement Swapping}

In classical communications, the message is transmitted sequentially across 
nodes, needing to reach repeater $k$ before being forwarded to repeater $k+1$. 
Quantum communication utilizes entanglement's advantage to overcome this 
constraint. In a quantum network, all entangled pairs are generated
simultaneously and  transmitted across each of the quantum channels.
Subsequently, Bell State Measurements (BSM) are executed on the photon 
received from the preceding node and the stored one for every quantum channel. 
Depending on the outcome of the measurement at each node, a specific quantum 
gate needs to be applied to  the final qubit. This value is sent classically 
and applied in the last node. The complete protocol is referred to as 
Entanglement Swapping, and ends up with the generation of a heralded Bell state
$\ket{\Phi^+} = (\ket{00} + \ket{11}/ \sqrt{2}$ between two nodes not 
directly connected. In this process, two quantum memory positions are used.
In this paper, entanglement swapping is supposed to be run sequentially
across adjacent links, and we assume that multiple overlapped entanglement
generation requests can exist at a given time in a QR. These requests can 
be served under two possible policies: oldest qubit first (OQF), or newest
qubit first (NQF).

\subsection{Fidelity}
\label{subsec:Cost}

Fidelity is a measure of the amount of similarity between two quantum states 
$\rho$ and $\sigma$. It is defined as
\begin{equation*}
     F(\rho, \sigma) = \text{Tr}\sqrt{\sqrt{\rho} \sigma \sqrt{\rho}},
\end{equation*}
where $\rho$ and $\sigma$ are density matrices. Fidelity is symmetric, has values 
$0 \leq F(\rho, \sigma) \leq 1$ and is maximal $F(\rho, \sigma) = 1$ whenever 
$\rho = \sigma$.  Thus, $F(\rho, \Omega)$, where 
$\Omega = \ket{\Phi^+}\bra{\Phi^+}$ is a maximally entangled state, can 
be interpreted as a  measure of the amount of entanglement of the state $\rho$.
For many scenarios of interest in quantum information
problems, the calculation or measurement of entanglement involves only pure 
states, and in this case the definition simplifies to
$F(\ket{\psi},\ket{\phi}) = \lvert \bra{\psi} \ket{\phi} \rvert^2$.
In this paper, fidelity will be taken as the metric for the quality of 
the entangled states created  between two endpoints in the network.

\subsection{Purification}

Purification is a quantum processing task wherein auxiliary quantum states
are used to improve the fidelity of a noisy entangled state, boosting its 
quality. This can be achieved in multiple forms, so we explain here only one,
the DEJMPS purification mechanism~\cite{Deutsch1996} as a representative. 
Assume a two qubit state in a node that belongs to the ensemble 
$\{ (\lambda_1, \ket{\phi^+}), (\lambda_2, \ket{\psi^+}),
(\lambda_3, \ket{\phi^-}), (\lambda_4, \ket{\psi^-}) \}$. Its
density matrix is
\begin{equation}
    \label{rhoent}
       \rho =  \lambda_{1} \Phi^+ + \lambda_2 \Psi^+ + \lambda_3 \Phi^- + \lambda_4 \Psi^-,
\end{equation}
where $\Phi^+ = \ket{\phi^+}\bra{\phi^+}$, and similarly for the other Bell 
states, which is trivially diagonal in the Bell state basis. The fidelity is
\begin{equation}
    F(\Phi^+, \rho) =  \bra{\phi ^+} \rho \ket{\phi ^+} = \lambda_1.
\end{equation}

Let us now suppose that Alice ($A$) and Bob ($B$) share two copies of 
the entangled pair~\eqref{rhoent}, $\{A_0,  B_0\}$ and  $\{A_1, B_1\}$. 
Firstly, they both need to decide which entangled pair they want to keep 
in memory, i.e., which of these is distilled. Without loss of generality, 
suppose that  they keep $\{A_0, B_0\}$. Now, $A$ and $B$ perform a $R_x$ 
rotation gate on each of the qubits, Alice applies $R_x(\pi/2)$ and 
Bob $R_x (-\pi / 2)$, where
\begin{equation}
   R_x(\theta) = \begin{pmatrix}
\cos\left(\frac{\theta}{2}\right) & -i\sin\left(\frac{\theta}{2}\right) \\
-i\sin\left(\frac{\theta}{2}\right) & \cos\left(\frac{\theta}{2}\right).
\end{pmatrix}
\end{equation}
In the third step, a $CNOT$ gate with qubits $(A_0, B_0)$ as control and 
the other as target is executed, and a measurement is applied on the 
target  qubits. The measurement results $m_A, m_B$ are shared, and if they 
are equal ($00$ or $11$) the purification is successful and 
the communication is resumed. Otherwise, both pairs are discarded.
The output state $\rho_{out}$ is of the form~\eqref{rhoent}, but with 
a higher values of $\lambda_1$ (i.e. fidelity), in fact with
\begin{equation}
    F' = \frac{\lambda_1^2+\lambda_4 ^2}{(\lambda_1+\lambda_4)^2+(\lambda_2+\lambda_3)^2}.
\end{equation}
The purification process is probabilistic, succeeding with probability
$(\lambda_1+\lambda_4)^2+(\lambda_2+\lambda_3)^2.$ It  is easy to check 
that $F' > F$ for $\lambda_1 > 1/2$.

In this paper, we will assume that the qubit that is sent is subject to 
depolarizing noise, so~\eqref{rhoent} is changed to the following
\begin{equation}
\label{rhopart}
    %\begin{split}
       \rho =  \left( 1- \frac{3p}{4} \right) \ket{\phi^+}\bra{\phi^+}+ \frac{p}{4} \ket{\psi^+}\bra{\psi ^+} +
        \frac{p}{4}  \ket{\phi ^-}\bra{\phi^-} + \frac{p}{4}  \ket{\psi^-}\bra{\psi ^-}.
    %\end{split}
\end{equation}
The condition $\lambda_1 > \lambda_2 > \lambda_3 > \lambda_4$ is 
trivially fulfilled $\forall p < 1$, and $\lambda_1 > 1/2$ is sustained 
for $p < 2/3$, where $p$ is the depolarizing probability of the Pauli 
channel.

\subsection{Memory and channel Depolarization}

Depolarization describes an application of isotropic noise to a quantum 
state, and can be interpreted as a uniform contraction of the Bloch sphere.
Depolarization is a form of time decoherence that can be modeled mathematically
as the quantum channel
\begin{equation}
    \label{eq:depolarizing-channel}
    \mathcal{D}_T(\rho) = e^{-T t} \rho + (1 - e^{-T t} \frac{\mathds{1}}{2}),    
\end{equation}
where $T$ is the decoherence decay rate and $I$ is the identity matrix. 
The depolarizing channel can be seen as a particular case of a single-qubit
Pauli channel, and its action of a maximally entangle state is described by
\begin{equation*}
    \mathds{1} \otimes \mathcal{D}_T(\Phi^+) = e^{-T t} \Phi^+ + (1 - e^{-T t} \frac{\mathds{1}}{4}).
\end{equation*}
Depolarization can also affect the qubits transmitted through a physical 
channel, namely photons in an optic fiber, and in this case this noisy
transmission can be represented by the formula
\begin{equation}
    p_{\text{depol}} = 1 - ( 1 - p_{\text{in}}) 10^{-L^2 \cdot \eta/10}
    \label{Eq:caidaLcuadrado}
\end{equation}

where $p_{\text{in}}$ denotes the probability of photon loss upon entering 
the fiber,  $L$ (Km) is the channel length, and $\eta$ (dB/Km²) is the
unit channel  attenuation. $p_{\text{depol}}$ is the probability that the 
qubit (the carrier photon) gets completely depolarized ($\mathds{1}{2}$)
and is useless for communication or computing.

In this paper, memory decoherence is modeled through a depolarization channel
that exponentially disturbs the quantum state $\rho$, as above
in~\eqref{eq:depolarizing-channel}, whereas channel depolarization exhibits a
probabilistic behavior which is not that of~\eqref{Eq:caidaLcuadrado}. We introduce our 
probabilistic model for depolarization in
Section~\ref{sec:implementation}.

\section{Model for Entanglement-based Networking}
\label{sec:Description}

\begin{figure}[pt]
\centering
\includegraphics[width=0.8\columnwidth]{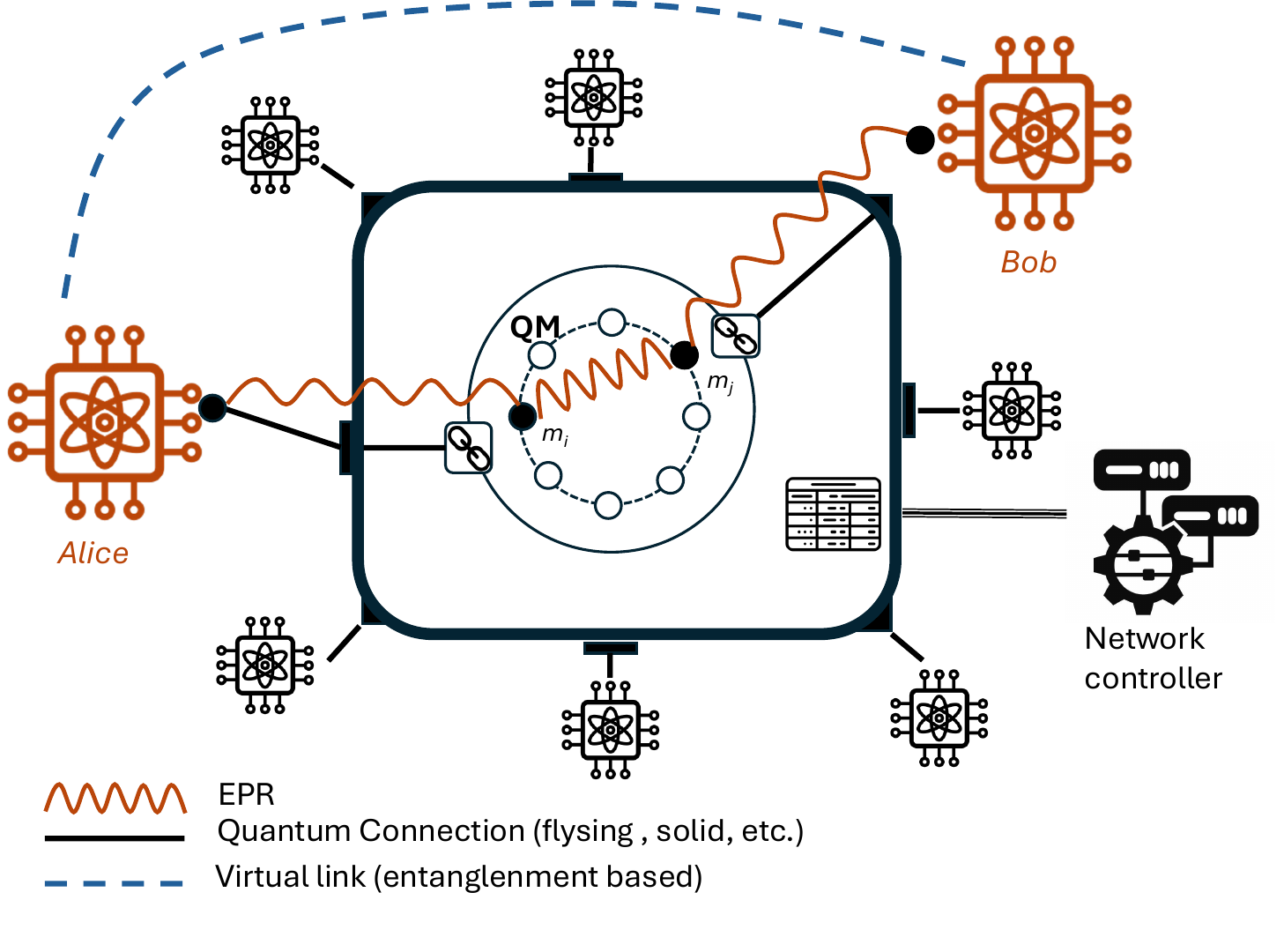}
\caption{\label{fig:qswitch} Intuition behind the quantum switch. Several nodes are displayed around it and three of them request a connection amongst them.}
\end{figure}   

We adopt in this work as the general model for a quantum network an abstract 
weighted directed graph $G = (V, E, \{ \omega_e \}_{e \in E})$ where the 
vertices $v \in V$ represent either end-nodes or intermediary nodes, the edges 
$e \in E$ represent communication links (classical and  quantum) between 
connected nodes, and each edge is labeled with a weight $\omega_e \geq 0$, 
which in our case will be the average fidelity between the entangled pairs 
created at $e$. Hence, we assume that any point-to-point link 
can be used for creating link-level entanglement  between two nodes, a 
shared state to be used for some specific communication ---quantum or 
entanglement-assisted--- or an arbitrary quantum computation task. An end node 
is just a source or sink of quantum information, i.e, any physical device 
capable of creating, operating on, or measuring qubits (more generally, 
$n$-dimensional quantum states). 

Intermediary modes are generic quantum switches  (see Figure~\ref{fig:qswitch}) which support the creation of end-to-end entanglement. A quantum switch is modeled as a 
device having $k$ input/output links and $m$ quantum memory units, where  each memory unit can store one qubit. 

The links can connect two adjacent intermediary nodes, or intermediary and  end-nodes, and 
switching switch generate extended entanglement between a subset of its connected users.  
This is accomplished by first generating link-level entanglement between the users and
intermediary nodes, followed by an entangling measurement on the quantum memories to 
create the  shared entangled state between the intended users (this process is called 
entanglement swapping (ES)~\cite{Vardoyan2021}). If a sequence of ESs 
are conducted on a path on the network graph $G$, the outcome will be an  end-to-end 
entanglement (EEE) association between a set of end users in the 
quantum network. This EEE bond will also be referred as a quantum path. Moreover, by 
assumption, multiple paths can simultaneously exist between an origin and a destination.  
s

From this graph model for Entanglement-Based Networks (EBN) where request from $s$ to $d$ can be served by  multiple paths, this paper focuses on linear topologies where quantum repeaters (QRs) are connected to two next nodes and support one single path. The main goal of this paper is to estimate (by simulation)  the feasible EEE fidelity on a linear topology of quantum repeaters. As in other works, we suppose that for generating link-level entanglement,  entangled qubit (Bell) pairs are continuously generated between the link endpoints, and that each  part these bipartite states can be temporarily stored in a single memory unit  until spontaneous decoherence destroys the entanglement. Upon an incoming  request, the QR selects a pair of its stored active qubits, performs a Bell measurement on these and generates an 
$2$-qubit entangled state $\ket{\Phi^+}$ between two of its links. 
Note that, in our 
model, the memory registers are functionally identical, finite in number, and 
have a deterministic finite decoherence time beyond which they are  useless for 
 EEE. Such coherence time emerges out of some physical property dependent
on the physical realization of the quantum register (see 
e.g.,~\cite{Huebel2007,Duer2007,Herbauts2013}),  and is a configurable 
parameter in  our setting. In contrast, the fidelity of a generated Bell 
pair, after a measurement, or after ES is probabilistic. Furthermore, instead 
of  considering that the latter operations succeed with a fixed probability 
(as in~\cite{Vardoyan2021}), our switch relies on physical probabilistic 
models for these key processes (see Section~\ref{sec:background}). Since, 
in particular, ES cannot guarantee an increase in the fidelity of the 
outcome bipartite state, we will allow the switch to  use entanglement
purification as a mechanism for boosting the fidelity to values  close  to 
$1$ (maximal entanglement) using some of  the known distillation
protocols~\cite{Deutsch1996}. 

\section{Simulation framework}
\label{sec:implementation}

\begin{figure}[t]
\centering
\includegraphics[width=0.9\columnwidth]{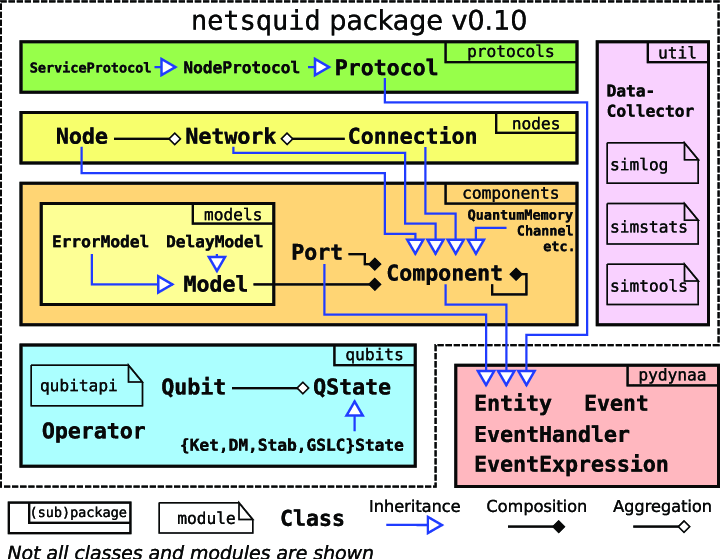}
\caption{Overview of software architecture in NetSquid (extracted from ~\cite{Coopmans2021}).
    \label{fig:NetSquid}}
\end{figure}   

According to our model  in Figure~\ref{fig:qswitch}, we have designed and developed a NetSquid simulation setting 
 for EDNs. NetSquid~\cite{Coopmans2021}  offers a range of foundational classes for the establishment of a quantum network (Figure \ref{fig:NetSquid}). When coupled with the discrete-event simulation engine PyDynAA, a comprehensive quantum network simulation can be executed. 

The lifecyle of  a simulation in  NestSquid consists on 3 steps: 
\begin{enumerate}
\item A network is constructed by using the subpackages \texttt{nodes} (main 
network elements) and \texttt{components} (physical elements \& models).
\item Protocols are assigned to network nodes to establish the networking 
behaviour.

\item The simulation is compiled in PyDynAA.
\end{enumerate}
NetSquid software is designed around PyDynAA entities which are capable of 
generating and receiving events (class \texttt{Event} in the core of PyDynAA) these events are automatically triggered by 
the simulation core and have a class type (\texttt{EventType}) which allows 
to construct event expressions to wait for (\texttt{EventExpression}). A NetSquid \texttt{protocol} 
is a virtual entity which describes the logic behind a quantum network. There
is not a clear notion of layers as in a classical networking architecture. 
However, a hierarchy can be set up in NetSquid by using protocols and 
sub-protocols, i.e., a network protocol running on multiple nodes can start 
sub-protocols that run on one node. Different protocols are coordinated by a
signaling mechanism to indicate a status change or announce a result. The 
physical behaviour of a component is described by the \texttt{Model} instances assigned to it,
which can specify various classical and quantum physical characteristics, such 
as transmission delays or noise.

Our  NetSquid-based simulation setting for EDN  uses standard classes in NetSquid for the channels,
i.e the combination of  \texttt{QuantumChannel} and
\texttt{ClassicalChannel} instances to provide a quantum connection between two nodes by 
using ports. From these standard connection elements in NetSquid, our EEE module implement a main class to deploy  
 the topology, \texttt{SetUp\_Network} and two specific node classes  for the logic of end-nodes and
switches in the network (\texttt{EndNode} and \texttt{Switch}). \texttt{Switch}
 inherits from the \texttt{Node} 
class and adds the attributes (memory and EPR sources) and methods for managing
the concurrency of entanglement swapping operations. On the other hand,
\texttt{EndNode} class has the logic associated with an end node. It inherits
from the \texttt{Node} class, and  adds the properties and methods for  managing the 
send buffer (queues) of qubits to be transmitted in the case of an Alice role. 
Finally, \texttt{SetUp\_Network} constructs the network by creating instances for connections, switches, 
and end-nodes.  

NetSquid protocols are the elements which include the dynamics in the simulation. 
Three  protocols are provided to support EEE in a quantum network: 
(1) \texttt{SwapProtocol} that waits for a qubit in the two associated memory
locations; performs a Bell measurement using the \texttt{SwapCorrectProgram} 
and sends the measurements to the end node; (2)   \texttt{CorrectProtocol},  
in the end node with Bob role,  that receives all the measurements and notifies
the coordination protocol; (3) \texttt{DistilProtocol} that contains the
distillation logic (purification) to be executed in the source and destination
nodes. %Also a routing strategy is needed to coordinate all the previous protocols and to establish an entanglement-based path (virtual link). To owe

In the NetSquid base code and snippets\cite{Coopmans2021, qrepchain, qswitch}, contributions have been aimed at concrete cases that support network architectures in which routing is not possible. Our contribution supports similar topologies with the addition of memory positions to add a layer of complexity and include E2E fidelity aware routing and purification \cite{zhao2022e2e}. Other numerical simulation tools, while offering more theoretical insights, tends to have a high computational load and is also limited to simplified quantum models. To the best of our knowledge, there seem to be no general non-analytical simulation solutions for entanglement-based quantum network topologies. In this paper we show our results in using the NetSquid-based EDN module to simulate topologies that interconnect a chain of QRs and multiple paths between end nodes.

\section{Case study: analysis of a repeater chain}
\label{sec:scenario}

\begin{figure}[t]
\centering
\includegraphics[width=\textwidth ]{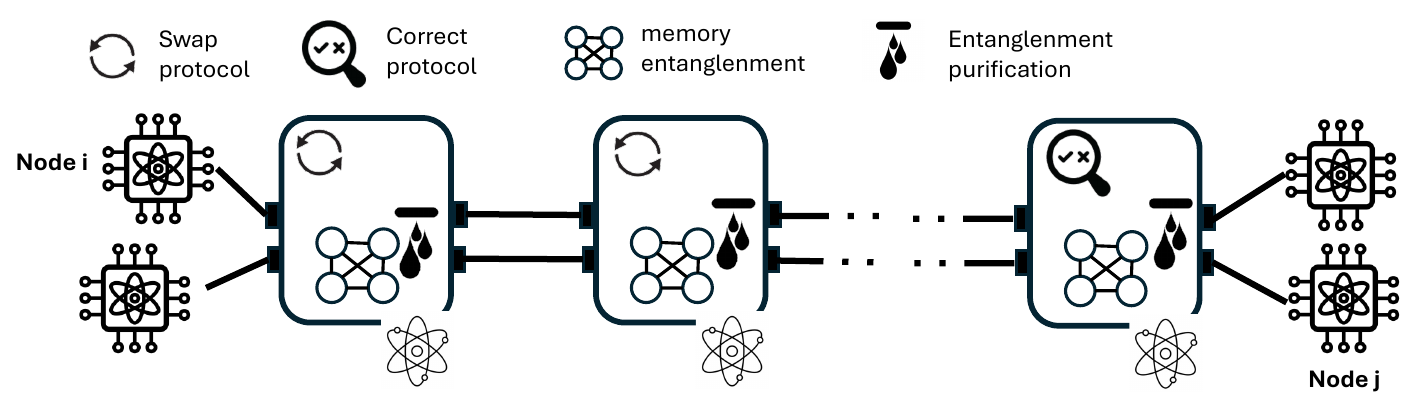}
\caption{Scheme for the quantum communication scenario, including  
the main quantum mechanism, swapping and correct protocols and purification.
    \label{fig:scenario-com}}
\end{figure} 
\begin{algorithm}[t]
\floatname{algorithm}{Algorithm}
\caption{Setup Network \label{NetworkManager}}
\small 
\begin{algorithmic}[1]
    \Function{setup\_network}{number\_of\_switches, distances, num\_positions}
    \State network = Network("linear\_network")
    \State EndNode end\_nodes[] 
    \State Switch switch\_nodes[]
    \State nodes=[end\_nodes[0]] + switch\_nodes + [end\_nodes[1]]
    \State network.add\_nodes(nodes)
    \For{i in range(len(nodes)-1)}
        \State left, right = nodes[i], nodes[i + 1]
        \State cconn = ClassicalConnection(distances[i])
        \State network.add\_connection(left, right, cconn)
        \For{j in range(num\_positions//2)} 
            \State qconn = QuantumConnection(j, distances[i])
            \State network.add\_connection(left, right, qconn)
            \State left.ports.forward\_in(left.qmemory.ports[])
            \State right.ports.forward\_in(right.qmemory.ports[])
        \EndFor
    \EndFor
    \State \textbf{return} network
    \EndFunction
    \end{algorithmic}
\end{algorithm}

As a case study for the programming environment described, we deployed 
the repeater quantum chain represented in the Figure~\ref{fig:scenario-com}, 
where intermediate switches are equipped with some memory positions and they 
are responsible to do  the swapping. Also, the final end node in the
communication, on the right, executes the measurements in the correct protocol. 
Additionally, purification may be deployed in intermediate and/or end nodes to 
improve fidelity.  Every QR contains  quantum and classical communication ports on 
the left and the right, and a  quantum program which executes gates in a 
quantum processor, mainly instruction-gates linked to the protocols and to 
the measurements. 

The topology is created  (Algorithm~\ref{NetworkManager}) from a network configuration file by adding end nodes and switches and then, for the linear topology in our case study,  by setting up the connections and linking the ports to the memory positions.   Then protocols are associated to nodes in the  master simulation script (Algorithm~\ref{simulation}), for our case study, a specific protocol (\texttt{repeaterProtocol} has been designed and developed. Apart from deploying quantum protocols over the topology, the script also collects (\texttt{dataCollector} performance and quality data, for instance, in order to apply a routing strategy based on VL availability and  end-to-end fidelity. 

\begin{algorithm}[t]
\floatname{algorithm}{Algorithm}
\caption{Simulation \label{simulation}}
\small 
\begin{algorithmic}[1]
    \Function{run\_simulation}{n, d, m} \Comment{n = num\_nodes, d = distances, m = mem\_positios}
       \State runtime = 1e9 \Comment{Arbitrary simulation time}
     \State network = setup\_network(n - 2,d,  ... ,m)
       \State rp = repeaterProtocol(network, mem\_pairs=m/2)
        \State dc = dataCollector(network, RepProtocol)
        \State rp.start()
        \State sim\_run(runtime)
        \State \textbf{return} dc.DataFrame
    \EndFunction
\end{algorithmic}
\end{algorithm}

\begin{algorithm}[t]
\floatname{algorithm}{Algorithm}
\caption{Repeater Protocol \label{RepeaterProtocol}}
\small 
\begin{algorithmic}[1]
 \Require SwapProtocol
    \Require CorrectProtocol
    \Function{repeater\_protocol}{network, mem\_pairs}
        \State  RepProtocol = LocalProtocol(network.nodes) \\
        // parent protocol comprising all nodes
        \For {node in network.nodes} \\
           \Comment{DistillProtocol can be included here, after or before swapping}
            \State RepProtocol.add\_subprotocol(SwapProtocol( node, ... ,mem\_pairs))
        \EndFor
        \State RepProtocol.add\_subprotocol(CorrectProtocol(lastnode))
        \State \textbf{return} protocol
    \EndFunction
    \\
    \Function{datacollector}{network, protocol}
        \Function {calc\_fidelity}{evexpr}
            \State qubit\_a, = nodes[0].qmemory.peek([final\_initialpos])
            \State qubit\_b, = nodes[-1].qmemory.peek([final\_endpos])
            \State fidelity =ns.qubits.fidelity([qubit\_b, qubit\_a], ks.b00, squared=True)
            \State \textbf{return} fidelity
        \EndFunction
        \State dc = DataCollector(calc\_fidelity)
        \State dc.collect\_on(protocol, SUCCESS)])
        \State \textbf{return} dc
    \EndFunction
  \end{algorithmic}
\end{algorithm}

For our chained topology, the repeater protocol is implemented as a protocol
that runs locally  in all the nodes of the topology, so that different 
sub-protocols can be launched in the single nodes.  The repeater protocol 
(Algorithm~\ref{RepeaterProtocol}) is responsible for launching  
sub-protocols for entanglement swapping (intermediate nodes) and the 
sub-protocol for applying corrections at the end-node (Bob on the right). 
With respect to the original NetSquid developments, we have deployed a 
scalable, configurable and extensible module. Although the 
original code structure is kept, our module allows to configure the  
memory positions available for communication, a class to register the 
fidelity and use that information to decide on which positions perform 
BSM and optimize resources, compatibility with purification protocols 
and a complete revamp of the noise model (depolarization).  
 
\begin{figure}[t]
\centering
\includegraphics[width=\textwidth ]{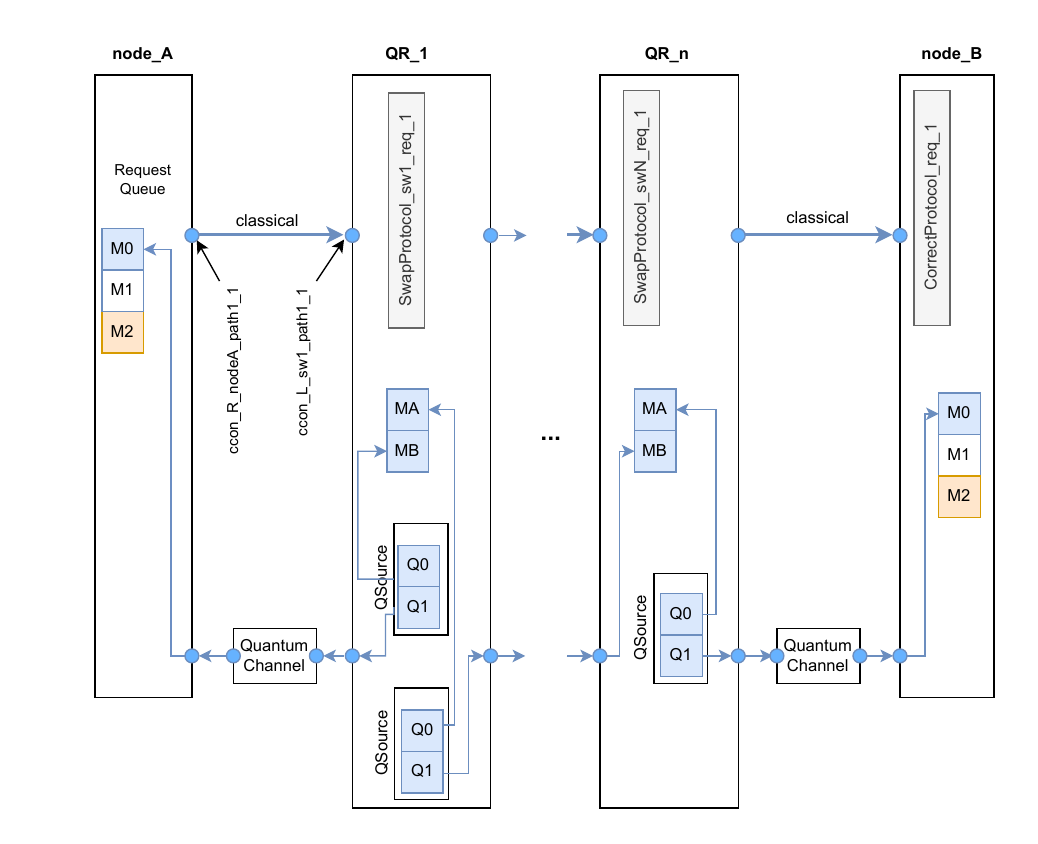}
\caption{Implementation architecture of the relay network in NetSquid \label{fig:swap-correct-protcol}}
\end{figure}

As an illustration, Fig.~\ref{fig:swap-correct-protcol} shows  for a given  virtual link, the software architecture of the protocols at the nodes (end nodes and repeaters QR\_1 to QR\_n) if purification is not required in the case of a repeater chain (not proper switches).  There is one instance of \texttt{SwapProtocol} at each switch traversed, and  one instance of \texttt{CorrectProtocol} at the destination end node. However, if purification is required, the number of links  is doubled, coordinating  two \texttt{SwapProtocol} at each switch, two \texttt{CorrectProtocol} at the  end node and the \texttt{DistilProtocol} at each source and destination end node. node\_A and node\_B are equipped with some quantum memory to store the pending requests (qubits to teleport from Alice to Bob). For each request of a virtual link between source and destination (node\_A and node\_B in the figure),  routing strategy will coordinate the other protocols to create the virtual link and, upon completion, it will notify the requester.

In the software architecture depicted in Fig. \ref{fig:swap-correct-protcol}, node node\_A and node\_B include a set of quantum memory positions to queue communications requests, these are the memory positions $M_0$ to $M_2$ for the information qubits. Also, every 
node $QR$  from $1$ to $n$ is equipped with two memory positions $M_A$ and 
$M_B$ and two EPR sources linked to these positions, one for the left and one 
for the right. The two memory positions $M_A$ and $M_B$ are used for swapping 
the qubits at the repeater. Although $QR_n$ is also equipped with 2 EPR sources,
juts one in needed to establish a unidirectional virtual link from source to
destination.   Classical links which assist to quantum communication are 
created according to the virtual path, in this case just one path for a fully
linear chain of repeaters. On the other hand, one instance of swapping and
correction protocols are executed for each communication request from Alice 
(node\_A) to Bob (node\_B), see the suffix request $1$ in the figure. 

We have used this module to simulate variable-length repeater-chains and
variable-capacity quantum memories at the repeater. The decision about how 
to overcome the physical restrictions to quantum communication should be 
a trade-off between the number of repeaters and the memory positions in 
the path.

\subsection{Multi-hop with 4 memory positions}
\label{sec:multihop}

In this scenario, QRs consist in four memory positions with labels from $0$ to $3$,
differing from the example architecture in Fig.~\ref{fig:swap-correct-protcol} with just two memory positions. By convention, even (odd) position numbers are employed to pair with the 
forward (backward) node. The motivation behind this structure is to have a
defined architecture for the memory positions, rather than a dynamic 
arrangement of the pairings, which is used to assess the fidelities of 
these pairs.  In this $3$-node portion of the network, only the middle node 
should perform Bell measurements on the states, as any subsequent attempt 
in obtaining the fidelity of already measured qubits fails. This redundancy 
is avoided by classically communicating the adequate positions to both 
forward and backward nodes. Another change we implemented is the improvement 
of the error model, because the default values were fundamentally flawed. 
Merging the depolarization probability with the attenuation of the fiber, 
although it may make sense from a computational standpoint, has no physical
value, as the depolarization is orders of magnitude lower than the
attenuation~\cite{Herbauts2013}.  Instead we use the polarization mode 
dispersion time $\tau_{\text{PMD}}(L)$, as a normally distributed variable 
with standard deviation $100\%$ to account for our depolarization probability. 
In such a model, we want to start to see depolarization effects when 
$\tau_{\texttt{PMD}} \geq \tau_{\texttt{coh}}=1.6$ ps. These 
results are thoroughly discussed in~\cite{Huebel2007}. For this, we 
employ the following formula
\begin{equation}
    p_{\texttt{depol}}=
   \begin{cases} 
      0 & \tau_{\text{PMD}} < \tau_{\text{coh}} \\
      1 & \tau_{\text{PMD}} \geq \tau_{\text{coh}}
   \end{cases}.  
\end{equation}

Lastly, for this physical configuration only, we changed the flow of the 
program to be able to compare two types of quantum architectures as
in~\cite{Panigrahy2022}, where the quantum protocols were interchanged: 
doing purification first on each link and then perform entanglement swapping 
on the nodes (PS) and first performing entanglement swapping on all the nodes
and then apply purification in the end nodes (SP). This differentiation will 
be used to check if the processes are interchangeable in terms of output
fidelity. With these improvements and the corresponding modifications, we 
can now properly simulate a network of $N$ nodes with two memory positions each 
and routing capabilities, based on fidelity. 

\subsection{2-hop  path with multiple memory positions} 
\label{sec:qswitch}

\begin{figure}[t]
\centering
\includegraphics[width=\textwidth ]{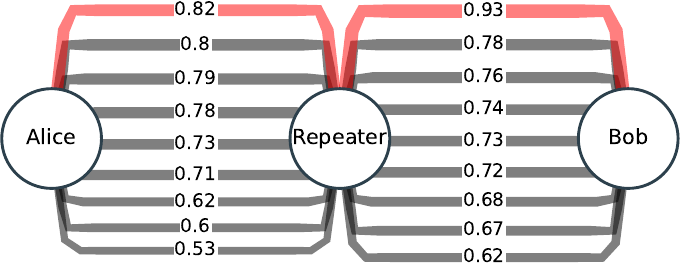}
\caption{\label{fig:QRepwithMemory}  Architecture discussed in Sec.~\ref{sec:qswitch}. 
A two-hop network is established and the program arranges and identifies the fidelity of 
the $m=18$ entangled pairs for each of the connections and the route with the highest 
fidelity is selected to drive the communication.}
\end{figure} 

This configuration benefits from all the changes introduced 
in~\ref{sec:multihop}, but does contain an arbitrary number $m$ of 
entangled memory positions between $3$ nodes, Alice, a switch, and Bob. 
This configuration is a sub-case of the network presented 
in~\cite{Vardoyan2021}, with two nodes instead of a generalized number 
of leaf nodes. An illustrative example of the behaviour of this network 
for $m=18$ can be seen in Fig.~\ref{fig:QRepwithMemory}. As we see in 
the Figure, the fidelity is increased 
the more pairs we introduce into the system, allowing us to use this overload 
of memories to drastically improve the communication capabilities of our 
system, on detriment of limiting the multiplexing capability of the device.

%\begin{figure}[h!]
%\begin{adjustwidth}{-\extralength}{0cm}
%\centering
%\includegraphics[width=2cm]{graph.pdf}
%\includegraphics[width=2cm]{graph.pdf}\\
%\end{adjustwidth}
%\caption{Main caption. \label{fig_label}}
%\end{figure}

\section{Results}
\label{sec:results}

\begin{figure}[tp]
\includegraphics[width=\textwidth ]{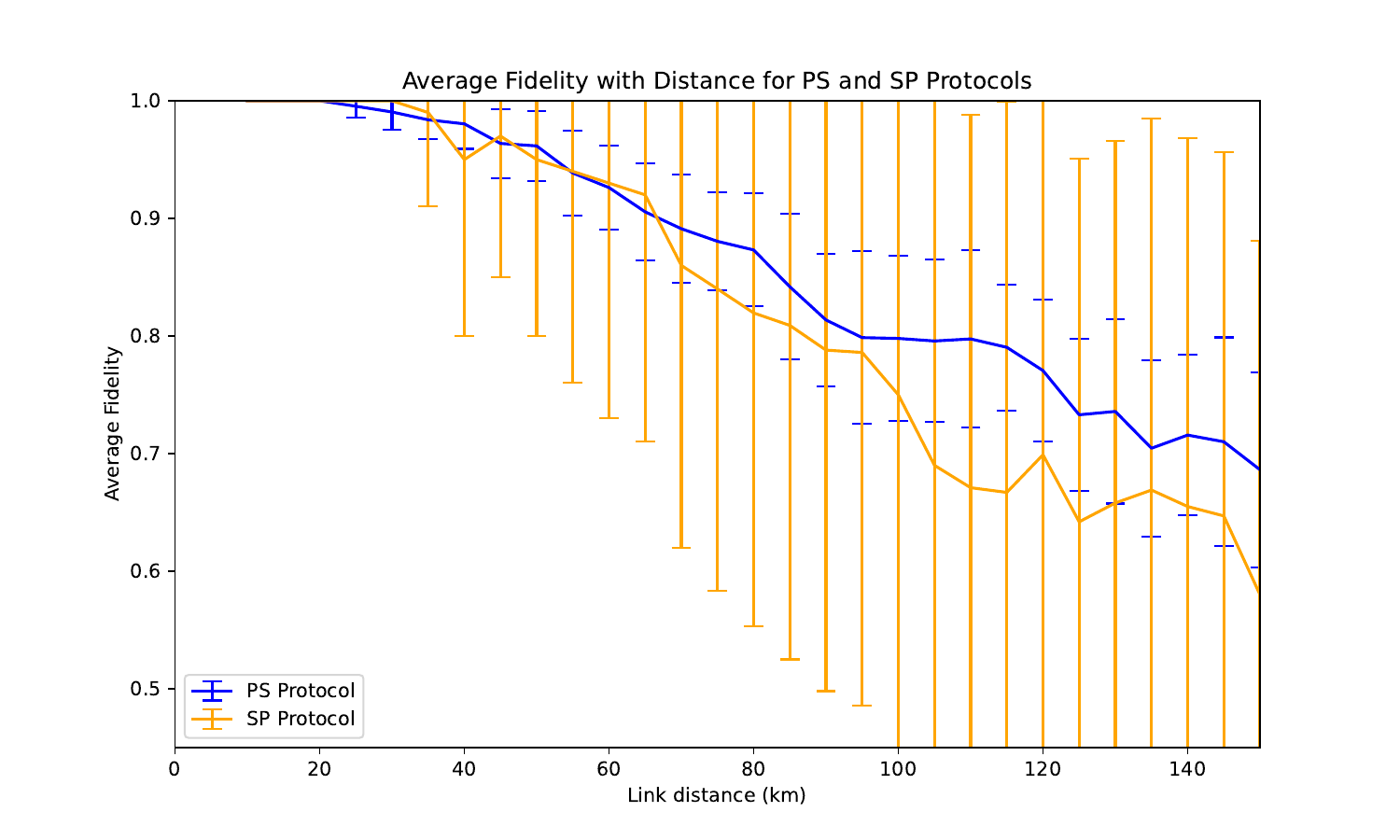}
\caption{(Color online) Fidelity average for the two different protocols discussed, PS and SP. These simulations have been performed in a network of 3 nodes. Error bars correspond to 1 standard deviation.
    \label{fig:sp_ps_comparison}}
\end{figure}     

\begin{comment}
\begin{figure}[tp]
\includegraphics[width=\textwidth]{depolreal.pdf}
\caption{Performance of different number of nodes with fixed distances 
    between end nodes. Nodes are equally distribute across the total distance. 
    Error bars represent the standard deviation of multiple simulations. \label{fig:firstsystem}}
\end{figure} 
\end{comment}

Firstly, we will compare two types of quantum architectures, PS
and SP~\cite{Duer2007}, as stated in~\ref{sec:multihop}. The former uses
purification before entanglement swapping (PS), whereas the opposite strategy
is used in SP. With the models proposed, the fidelity values between the 
architectures were similar for distances up to $70$ km, as can  be seen in Fig. \ref{fig:sp_ps_comparison}. At that point the 
PS protocol becomes better at the cost of less probability to 
achieve a successful purification (more time consumed). Because of this 
fidelity increase, we will proceed with performing purification of the 
qubits and then the swapping operations (PS) with correction gates applied at the end node qubit.

\begin{figure}[t]
%\centering
    \subfigure[Fidelity by varying the number of QRs (4 memory positions)  equally distribute across $L = 1000$ Km. \label{fig:firstsystem}]{\includegraphics[width=0.49\textwidth ]{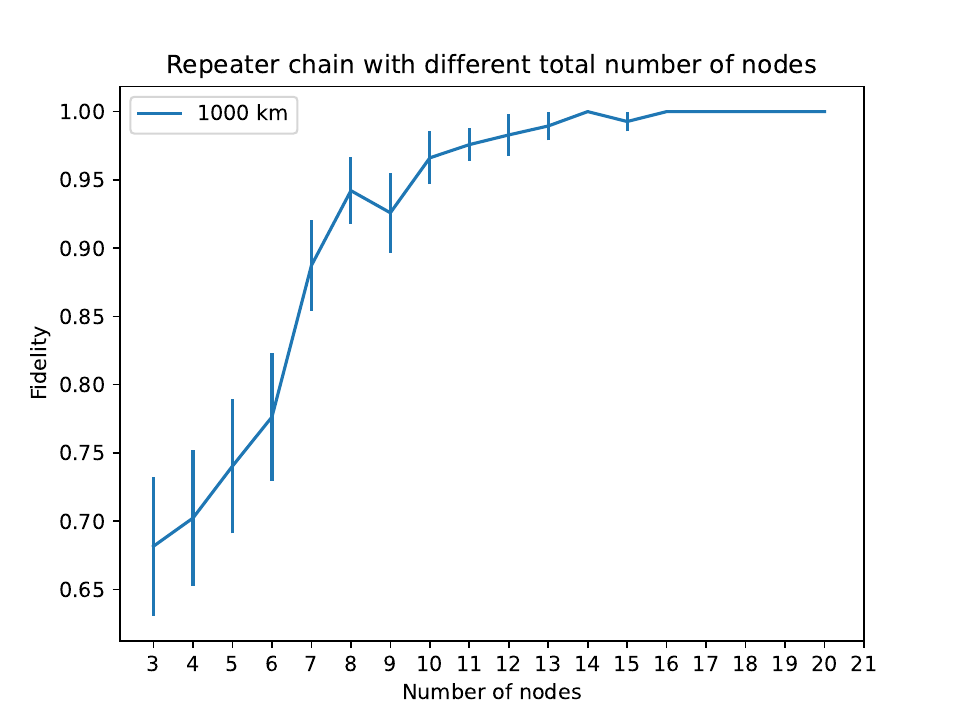}
    }
    \hspace{0.2cm}
    \subfigure[Fidelity by varying the number of memory positions in  $2$-hops with one QR and a inter-node distance $500$ km.\label{fig:secondsystem}]{
    \includegraphics[width=0.49\textwidth ]{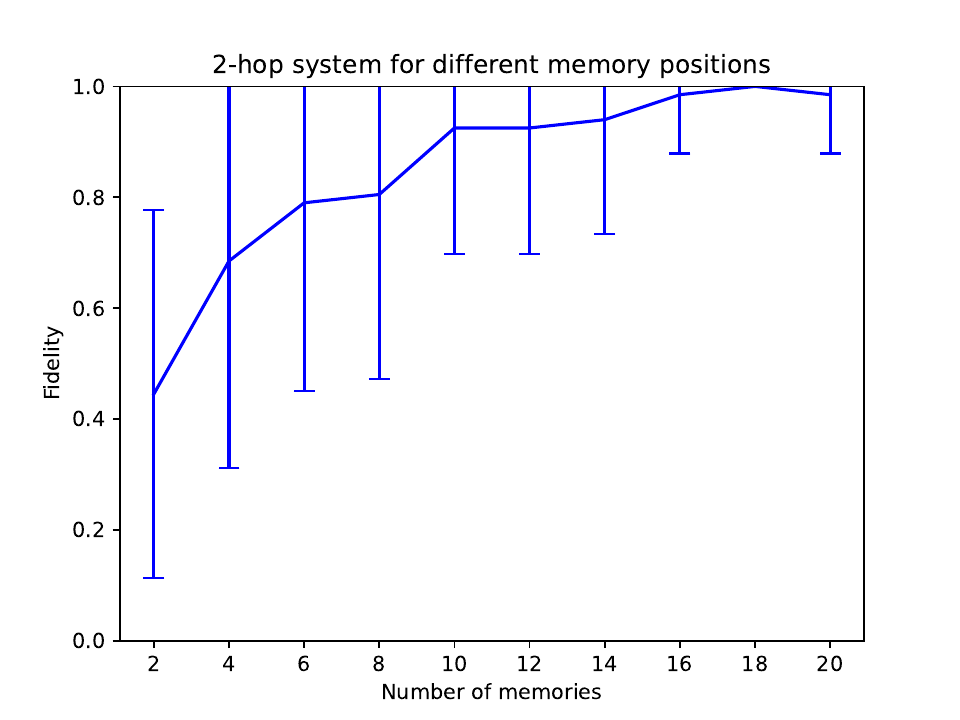}
    }
    \caption{Comparison of the impact on fidelity by varying the number of intermediate QR  (Fig.\ref{fig:firstsystem}) and by varying the number of memory positions in a QR  (Fig.\ref{fig:secondsystem}). Error bars represent the standard deviation of multiple simulations.}
\end{figure}

We implemented the simulation scenarios in sections \ref{sec:multihop}
(Multi-hop with fixed number of memory positions) and \ref{sec:qswitch} (2-hop  path with variable number of  memory positions) to determine the effectiveness of the system and establish the 
minimal resources needed to achieve practical quantum communication. Both scenarios are observed in the presence of depolarization error in quantum channels, which allows to study the influence of the number of repeaters and memory positions; other error models will produce similar results but with shorter distances. For instance, in Fig.~\ref{fig:firstsystem}, we evaluate the network
architecture discussed in Section~\ref{sec:multihop}, for a network of
total distance $L = 1000$ km, $4$ memory positions for each node, and a
source delay of $1$ ns. All sources of error have been suppressed except for channel depolarization, to give a better understanding of the implemented depolarization model. In this case, without purification, we see an inflection point at around 10 nodes, which correspond to an internode distance of 125 km, which should be the upper limit for quantum channels in terms of fidelity, as longer channels would rapidly decrease the fidelity, out of bounds for proper purification. Additionally, close to the same 10 nodes mark, there is the fidelity value of $F=0.95$, which the authors consider to be the minimum usable fidelity for applications, specially without purification.%For instance, in Fig.~\ref{fig:firstsystem}, we evaluate the network architecture discussed in Section~\ref{sec:multihop}, for a network of total distance $L = 1000$ km, $4$ memory positions for each node, and a source delay of $1$ ns. Memory noise, gate noise, and source fidelity are all considered ideal, only the channels present depolarization. Of course, the inter-channel distance is the only factor affecting fidelity when error only comes from channel depolarization. In this case, to achieve usable fidelity $(F > 0.95)$ we should not surpass the $100$ km threshold. This could  be overcome by the use of purification. 

\begin{figure}[H]
\includegraphics[width=\textwidth ]{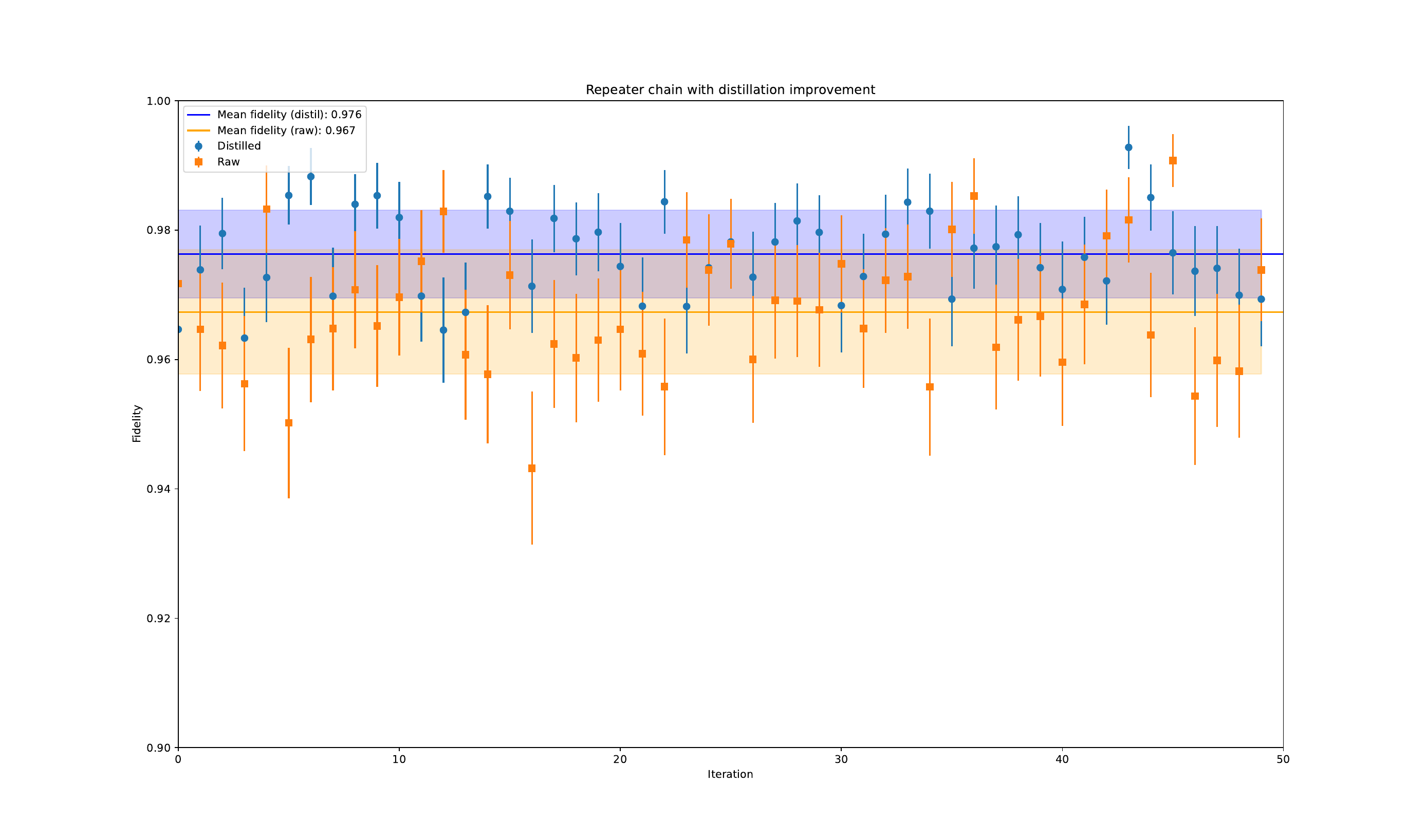}
\caption{Simulation for a three node network applying one round of DEJMPS
    distillation protocol.  \label{fig:distil 1 round}}
\end{figure}  

On the other hand, for the scenario in Section \ref{sec:qswitch}, we can  evaluate how fidelity behaves when varying the amount of memory positions. In  Fig.~\ref{fig:secondsystem}, we can see for a network of $1000$ km how 
the fidelity increases with the number of memory positions.
In a real scenario, attenuation of the signals would be the primary error source and this behaviour could not be appreciated, but we anted to focus on the depolarization model for these experiments.

\begin{comment}
\begin{figure}[tp]
\includegraphics[width=\textwidth]{memoriesfor2hop.pdf}
\caption{Fidelity for different amounts of memory positions in a 2-hop 
    network of inter-node distance $500$ km.  \label{fig:secondsystem}}
\end{figure}   
\end{comment}

As we can see in Fig.~\ref{fig:distil 1 round} we expect an improvement in 
the high percentile of the fidelity regions. This concludes that the 
theoretical advantage of the DEJMPS is present as well in the simulations,
although it consists of a rough $1\%$, the protocol has only been applied once. We now can compare these results with the increasing the number of memory positions within a node (Fig.~\ref{fig:secondsystem}) which increments the chances 
of establishing high fidelity links between the nodes by simply incrementing  the generated pairs per round. On the contrary, one sround DEJMPS (Fig.~\ref{fig:distil 1 round})
only uses one extra EPR pair in order to increase fidelity, although $R_X(\theta)$ and CNOT gates are involved, as well as a measurement, which are very resource intensive, specially in time.

\begin{figure}[tp]
    \centering
    \includegraphics[width=\linewidth ]{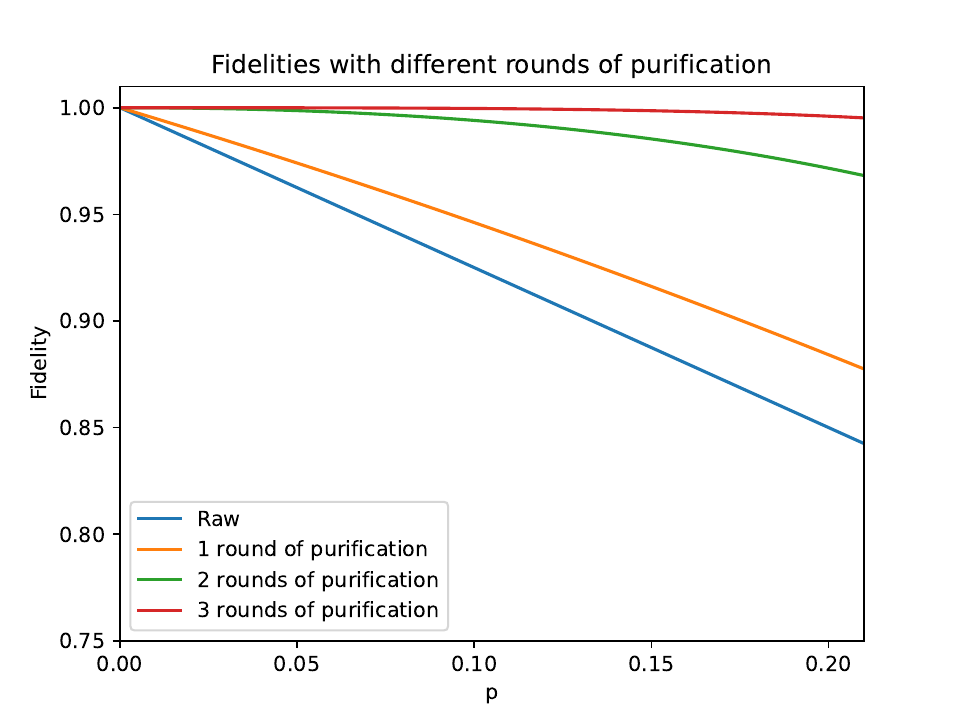}
    \caption{Representation of the theoretical increment in the fidelity after 
    applying several rounds of purification. The total depolarizing probability 
    of the channel is represented in the $x-$axis vs. the fidelity output of the 
    process.}
    \label{fig:many purification rounds}
\end{figure}

Furthermore, there is an intrinsic probabilistic nature associated to this  method, as both measurements registered by the two parties must coincide in  order to ensure a successful purification, as well as an asymptotic behaviour  that limits the effectiveness of both high fidelity inputs and an increment on  the number of rounds. As we see in Fig.~\ref{fig:many purification rounds}, the fidelity largely
increases after a second distillation round, whereas with a third iteration  the asymptotic behavior of the method enters into place and decreases the  gain in the parameter. This only strengthens the argument in favour of the application of DEJMPS distillation in order to increase the fidelity to
feasible limits for a successful communication, as the application of two  rounds of distillation only consume two additional EPR pairs, whereas if we were to apply the method in Fig.~\ref{fig:secondsystem} each of the nodes  would consume between $12-14$ memory slots to land into the higher 
percentiles of the fidelity.

Circling back to Fig.~\ref{fig:many purification rounds} we can extract 
an intuitive threshold for the tolerance of the fidelity around $F_{thr} 
\approx ~0.85$, as two rounds of purification take it close to $F > 0.95$,
which, for small and medium messages is a tolerable value of the parameter. 
This threshold needs to be changed if we are dealing with larger quantity 
of data and should be moved towards $F_{thr} \approx 0.90$. However, taking 
into account the asymptotic nature of the protocol a more complex fine 
tuning must be implemented in the communication scheme in order to transmit
large packages of data.

\section{Conclusions}
\label{sec:conclusions}

In this paper, we have argued that simulation is an essential tool for
understanding and designing quantum networks ---especially, networks based on
entanglement-distribution--- not only as a complement to simplified analytical
models, but mainly since entanglement is a physical resource inherently
different to classical network entities, like conventional protocols or 
switching devices. Moreover, the efficient generation and distribution
of entanglement prompts for a thorough rethinking of the internal 
architectures of quantum repeaters and switches. These architectural 
alternatives can be also addressed with a good simulation environment.

We have set the ground base model using NetSquid for quantum switches and quantum repeaters in general topologies, capable of handling an arbitrary
number of quantum memory units, taking fidelity as the primary
decision variable for path selection, and with depolarization noise
in the channels and decoherence in the memory registers. Using this programming environment,
fast prototyping and comparison between alternative network designs can be 
explored. As an example, we compared the advantages and performance of two 
typical configurations, a 2-hop network with multiple memory positions at
the QR versus a standard repeater chain, and determined the achievable end-to-end
fidelity in each case. This calculation is virtually intractable with
a purely analytical approach. For future work, a more general simulator will be implemented, both from the engineering standpoint, by supporting more complex topologies and protocols, and from the physics standpoint, by adding more control over the experimental parameters that define the network, and exploring different technologies for quantum hardware.

%\acknowledgments{TED2021-130369B-C31, TED2021-130369BC32, TED2021-130369B-C33 funded by MCIN/AEI/10.13039/501100011033 and by the “European Union NextGenerationEU/PRTR”. The work is also funded by the Plan Complementario de Comunicaciones Cu\'anticas,  Spanish Ministry of Science and Innovation(MICINN), Plan de Recuperación  NextGenerationEU de la Unión Europea (PRTR-C17.I1, CITIC Ref. 305.2022), and Regional  Government of Galicia (Agencia Gallega de Innovación, GAIN, CITIC Ref. 306.2022).3). Additionally, This paper has been also funded by the Galician Regional Government under project  ED431B 2024/41 (GPC).}

\bibliographystyle{ieeetran}

\end{document}